# Charge density wave ordering in NbSe$_3$: possible models and the experimental evidence

**A Prodan[1], H. J. P. van Midden[1], R Žitko[1], E Zupanič[1], J C Bennett[2] and H Böhm[3]**

[1]Jožef Stefan Institute, Ljubljana, Slovenia
[2]Department of Physics, Acadia University, Wolfville, Nova Scotia, Canada B0P 1X0
[3]Geosciences, University of Mainz, D-55099 Mainz, Germany

E-mail: albert.prodan@ijs.si

**Abstract**. Charge density wave (CDW) ordering in the prototypical low-dimensional compound NbSe$_3$ is reconsidered. We show that the widely accepted CDW model with two incommensurate modulations, $q_1$ = (0,0.241,0) and $q_2$ = (0.5,0.260,0.5), localized on type-III and type-I bi-capped trigonal prismatic (BCTP) columns, does not explain some details, revealed by various microscopic methods. The suggested alternative explanation is in a better accord with the entire experimental evidence, including low-temperature (LT) scanning tunneling microscopy (STM) results. It is based on the existence of modulated layered nano-domains formed below both CDW onset temperatures. According to this model, two of the three slightly different BCTP types of columns are modulated by the same wave vector, either $q_1$ or $q_2$, which can easily switch over in a domain as a whole. This approach explains the presence of the $q_2$ modulation in the STM images recorded above the T$_2$ CDW transition and the absence of the $q_2$ satellites in the corresponding diffraction patterns. The long periodic modulation, detected by LT STM is attributed to a beating between the two CDWs, centered on adjacent columns of the same type. These pairs of columns, both either of type-III or type-I, modulated by the two alternative CDWs, represent the basic modulation units, ordered into nano-domains.

## 1. Introduction

While charge density waves (CDW) are observed in a large number of low-dimensional systems, only a relatively small group of one-dimensional (1-D) compounds exhibits the phenomenon of CDW sliding under the application of an external electric field. Members of this group of compounds include in addition to NbSe$_3$ its isostructural monoclinic polymorph m-TaS$_3$ [1], NbS$_3$ [2], (TaSe$_4$)I [3], (NbSe$_4$)$_{10}$I$_3$ [4], the "blue bronzes" A$_{0.3}$MoO$_3$ with A = K, Rb, Tl [5-8] and the organic conductor TTF-TCNQ [9]. This rather limited list of compounds characterized by the extraordinary transport properties triggered a large interest and a detailed consideration of the phenomenon. By far the most thoroughly studied of the mentioned compounds was NbSe$_3$.

It was pointed out recently [10] that x-ray crystallography breaks down for structures in which order extends over nanometers only. NbSe$_3$ is a typical example where the extent of order in its modulated structure may be of crucial importance for its physical properties. We thus reconsider in the present work the available experimental evidence, particularly the recent low-temperature (LT) scanning tunneling microscopic (STM) results and try to give a concurrent explanation for the CDW ordering in this compound, which will hopefully be in a better accord with the entire available experimental evidence.

## 2. The basic structure

The NbSe$_3$ room-temperature (RT) basic structure [11] is shown in figure 1. It is constructed from three types of symmetry-related pairs of bi-capped trigonal prismatic (BCTP) columns. These columns are formed of Nb chains in Se cages, aligned parallel to the monoclinic $b_0$-direction. The inter-column covalent bonding forms corrugated layers parallel to the $b_0$-$c_0$ monoclinic plane, separated by van der Waals (vdW) gaps. Thus, the structure is strongly anisotropic in all three dimensions: in addition to the 1-D nature characterized by the BCTP columns, it also shows a pronounced two-dimensional character, whose origin is in the covalently bonded layers. Among the three different types of BCTP columns in NbSe$_3$ the type-I and type-III columns are rather similar, with one short Se-Se distance forming the equilateral bases of the prisms, while the bridging type-II columns appear more regular, with almost isosceles-like bases.

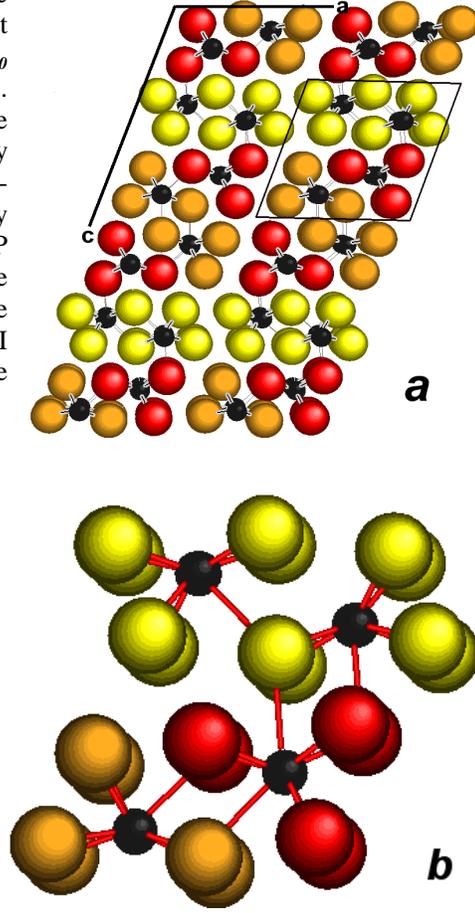

Figure 1 (Color online) The structure of NbSe$_3$ [11]. The red (dark), orange (gray) and yellow (light) balls represent Se atoms along type-II, type-I and type-III bi-capped trigonal prismatic cages with the 8-coordinated small black (dark) balls representing Nb atoms. (b) The marked detail with the shortest Nb-Se bonds indicated by red (dark) lines.

## 3. The CDW models

It was argued from the very beginning [12] that two incommensurate (IC) CDWs appear independently at different onset temperatures along two of the three available BCTP columns: one with wave vector $q_1 = (0,0.241,0)$ below $T_1 = 144$ K appearing along type-III columns and a second with wave vector $q_2 = (0.5,0.260,0.5)$ below $T_2 = 59$ K appearing along type-I columns. This generally accepted perception of CDW formation and ordering appeared to be in good accord with a variety of experiments performed on NbSe$_3$: in addition to the nonlinear transport properties attributed to CDW sliding and pinning [13-27], there were also electron [28] and x-ray diffraction studies [29-32], nuclear magnetic resonance (NMR) [33-35] and angular resolved photoemission spectroscopy [36,37] performed. However, the model seems to be in disaccord with some transmission electron microscopy (TEM) results [38] and particularly with the latest LT STM observations [39-42]. These difficulties are not to be ignored, not only because of the exceptional resolution achieved in the LT STM experiments, but also because of the very localized structural information this method is able to supply.

It was suggested later in an alternative model [43] that both IC CDWs are formed simultaneously and statistically along type-I columns below $T_1$ and in addition along type-III columns below $T_2$. According to this model the two CDWs are supposed to be separated into highly anisotropic and unstable layered nano-domains, oriented parallel to the vdW gaps. This model is based on two conditions: first, the domains must be sufficiently small to extinguish in the diffraction experiments the $q_2$ contribution to the reciprocal space between $T_1$ and $T_2$ and second, the domains must be in proper phase relationships. The first condition requires short correlation lengths of both domain types in comparison with the instrumental coherence regions, while the second requirement depends on the absolute values of the two $q$-vectors.

## 4. The experimental evidence

The available models for CDW ordering in NbSe$_3$ are to be evaluated on the basis of the following relevant experimental evidence.

*X-ray and electron diffraction*: Tsutsumi et al. [28] first reported observation of diffuse precursor scattering in electron diffraction patterns of NbSe$_3$ at temperatures above the $q_1$ CDW transition. Next, several x-ray analyses were performed on both, the $q_1$ and $q_2$ satellites. By measuring the line widths of both types of CDW satellites Fleming, Moncton and McWhan [29] found an anisotropy ratio of an order of 5 between the correlation lengths $\xi_{b*}$ and $\xi_{a*}$, i.e. parallel to the chains and perpendicular to the cleavage planes. The estimated parallel values a few degrees above T$_1$ and T$_2$ were $\xi_{b*} > 5$ nm (measured on a $q_1$ satellite at 160 K) and $\xi_{b*} > 10$ nm (measured on a $q_2$ satellite at 62 K), while the corresponding perpendicular values were $\xi_{a*} \approx 1.2$ nm and $\xi_{a*} \approx 3$ nm. Moudden et al. [30] determined after proper deconvolutions of the scattering profiles for the $q_1$ satellites a resolution limited correlation length $\xi_{b*} > 280$ nm at 80 K (i.e. bellow T$_1$) and $\xi_{b*} \approx 70$ nm at 155 K for scans along $b_0$* and $\xi_{a*} \approx 4.2$ nm, $\xi_{b*} \approx 14$ nm and $\xi_{c*} \approx 0.5$ nm for the transverse scans along $a_0$* at about 6 K above T$_1$. They also reported a slight variation of the $q_1$ vector between T$_1$ and 80 K. Rousière et al. [31] performed x-ray measurements of the pretransitional structural fluctuations for both, $q_1$ and $q_2$ satellites. They found at 72 K a smaller anisotropy for the $q_2$ satellites (1:3.5:6) as compared to the one of the $q_1$ satellites (1:4:20), with the ratios given along the perpendicular $b_0$, $a_0$, and $c_0$* directions, respectively. The intrachain correlation lengths $\xi_b$ and the interchain $\xi_a$ values were found to be a few degrees above the transition temperatures for both CDW wave vectors similar, while the corresponding $\xi_{c*}$ values were at larger variance. Approximate values for the $q_2$ satellites at 65 K (deduced from [31]) are $\xi_b \approx 6.5$ nm, $\xi_a \approx 2$ nm and $\xi_{c*} \approx 1.5$ nm. Such correlation lengths, particularly those measured perpendicular to the chains, indeed appear short in comparison with the coherence regions, whose estimated sizes for synchrotron x-rays radiation extend 125 nm [44] and for high-energy electrons in conventional TEM experiments 380 nm [45].

A structural analysis of the modulated structure was performed with synchrotron radiation [32]. The atomic displacements along the three BCTP columns were found to be in accord with the original Wilson´s model. However, if segments of variable lengths along the type-III (below T$_1$) and in addition along the type-I columns (below T$_2$) are modulated alternatively by the $q_1$ and $q_2$ wave vectors, as suggested in the domain model, the overall contribution to the reciprocal space would in fact appear identical for both suggested models. The same arguments also apply for any high energy electron diffraction experiments. Thus, if a disorder on a nanoscale indeed takes place in NbSe$_3$, both models cannot be distinguished on the basis of the corresponding diffraction experiments.

*NMR*: NMR measurements were performed at room temperature (RT), 77 K (LNT) and 4.2 K (LHeT) on powdered samples and on samples made of a large number of properly aligned crystals [33-35]. The $^{93}$Nb (I=9/2) spectra recorded at RT with the magnetic field parallel to the crystallographic $b_0$ direction resolved 27 lines, which clearly corresponded to the three different Nb sites. While one set of lines remained unchanged on cooling to the LHeT, one of the remaining two sets of lines became smeared at LNT and the remaining set in addition at LHeT. This behavior is indeed in accord with the original model [12], where the Nb sites of the type-II columns are not supposed to be affected by the CDW formation, while the $q_1$ CDW is expected to appear on cooling below T$_1$ along the type-III columns and in addition the $q_2$ CDW below T$_2$ along the remaining type-I columns. However, the behavior is also in accord with the domain model [43]; the problem with NMR is that the IC nature of the two CDWs clearly reveals contributions from different types of columns, but does not distinguish the corresponding Nb sites with regard to $q_1$ and $q_2$.

*TEM:* In addition to the early high-energy electron diffraction studies [28] NbSe$_3$ was also investigated by satellite dark field TEM [38]. Elongated strands, in average 20 nm wide and 2000 nm long, crossed by unstable Moiré-like fringes with spacings from only 8 nm to a few 100 nm, were observed in samples cooled to temperatures both above and below T$_2$. The strands and fringes were observed in addition to other contrast effects, which altogether exhibited a characteristic unstable "twinkling". The appearance of these features above T$_2$ could not be properly explained on basis of the original Wilson's model. Thus, the described TEM experiments were the first indication that the observed instabilities might be inconsistent with the generally accepted model.

***LT STM***: In spite of their high resolution the recently published LT STM images of NbSe$_3$ [40-42] are difficult to interpret. The main problem represents the variation of the images with the tunneling parameters, which vary the detected CDWs, particularly their intensities. In addition, subsurface effects are regularly detected superimposed onto the surface contribution, which complicates the interpretation. Nevertheless, there are a few details, some in accord with both models (points 1. to 3.), while the rest seem to support a modified domain model:

1. Dependent on temperature, all images shown reveal either only one (type-III) or two (type-III and type-I) strongly modulated BCTP columns along the basic structure unit cell periodicity *c$_0$*.

2. The intensity of the CDWs recorded varies largely with the tunneling parameters, particularly with a switched polarity of the applied gap voltage.

3. The strongly modulated columns are along the *c$_0$* direction ordered either in-phase (if there is only one strongly modulated column per *c$_0$*) or out-of-phase (if these appear in pairs). In the second case the adjacent pairs as units are also ordered out-of-phase, which automatically enlarges the lateral periodicity into 2*c$_0$*.

4. All strongly modulated surface columns show within the well ordered regions the same CDW periodicity along the columns.

5. Regardless of the actual labeling of the columns, there is always one mode present (either as a strong surface modulation or as a superimposed weak contribution from bellow the surface), whose ordering enlarges the lateral periodicity into 2*c$_0$*.

While the observation described under point 4. clearly supports the suggested domain model, the one described under point 5. raises doubts about this model as well. In this context it is also interesting to reinvestigate the first of the published LT STM studies on NbSe$_3$, not only because it triggered at the time it was published a discussion about its interpretation [39], but also because it shows the best resolution and thus supplies some key evidence for the evaluation of the models. The image reveals pairs of strongly modulated chains along *c$_0$*. These are both modulated by the same CDW and ordered out-of-phase, as described under point 3. above. With the lateral periodicity enlarged into 2*c$_0$*, the two adjacent columns forming pairs can only represent type-III and type-I columns, both modulated by the same *q$_2$* CDW. This is clearly in support of the domain model. What seems to be in disaccord with it are the remaining single weakly modulated columns. According to their position and intensity they obviously represent subsurface columns, which form pairs with the adjacent strongly modulated surface columns. The modulation periodicity along these columns appears slightly larger as compared to the one along the strongly modulated surface columns with an in-phase ordering along *c$_0$*. Thus, they seem to represent the alternative *q$_1$* CDWs. According to their position along *c$_0$* the adjacent surface-subsurface pairs can only represent the type-I pairs of columns; the type-III subsurface columns are positioned bellow their surface counterparts and cannot be detected. All columns show in addition to the modulation also an atomic resolution, however only one chain per column. This is in accord with expectations in case of the type-III and type-I columns, whose trigonal prismatic bases are equilateral with one short Se-Se bond, but is somewhat surprising in case of the remaining unmodulated chains. In accord with their position these show the surface type-II columns, whose trigonal prismatic bases are almost isosceles. Although these details are in accord with the basic supposition of the domain model, i.e. the existence of layered domains with both type-III and type-I columns modulated by either of the two CDWs, they are in disaccord with its prediction that the same CDWs appear along strongly bonded pairs of column. The described details rather indicate that the strongly bonded layers are composed of *q$_1$* and *q$_2$* sublayers, whose interchanging results in the domain structure.

## 5. Discussion

LT STM is of particular importance, because the method is basically different from the alternative methods, such as NMR, electron spin resonance, various diffraction methods, and last but not least even high-resolution TEM. Although these approaches are beyond any doubt capable of achieving resolution on an atomic scale, the scanning probe microscopies, and STM in particular, are the only methods where information is not collected statistically over a relatively large sample. STM has also drawbacks; it is a surface sensitive method and the interaction between the scanning tip and the surface may be sufficiently strong to induce changes in the surface. But in spite of that, it is certainly capable of revealing details like nano-domains, which might due to their small sizes and averaging within experimental coherence regions be overlooked by all methods based on statistics, regardless of their high resolution. This by no means reduces the importance of the alternative methods, but rather requires a comparison of all results, which should together lead to a single acceptable conclusion.

Next, there is the question of the STM image formation and its interpretation. Due to the Fermi surface shape, the major contribution to the surface density of states in NbSe$_3$ is supposed to come from the corrugated top Se layer [41]. However, it should be taken into account that a charge distribution can also be detected over larger distances and may show details located below the surface. Such effects, attributed to a charge transfer to the surface have been observed before, e.g. in the superstructures formed by intercalated metal atoms in the subsurface vdW gaps of NbS$_2$ and NbSe$_2$ [46] and in the case of unstable domain boundaries formed along such gaps in the monoclinic NbTe$_2$ [47]. The observed long-range ordering in NbSe$_3$ [41] must be of a similar origin, as schematically shown in figure 2a. A periodicity of 116 $b_0$ is chosen along both columns, with the first half (i.e. 58 $b_0$, the LP that accommodates 14 $q_1$ and 15 $q_2$ CDW modulation periods [43]) of the top column modulated with $q_2$ and the second half with $q_1$, while the column beneath is modulated in the opposite sense. If the charge modulations along two of such adjacent columns are indeed detected together, characteristic beating should be observed in the STM images. If in addition the $q_2$ sections are statistically displaced by half shorter $q_1$ sections (7 $q_1$ periods extending over 29 $b_0$ only [30]), as shown in figure 2b, the close proximity of the two columns would result in a $q_1$ contribution to the reciprocal space only [43]. However, the $q_2$ sections would still be detected by STM.

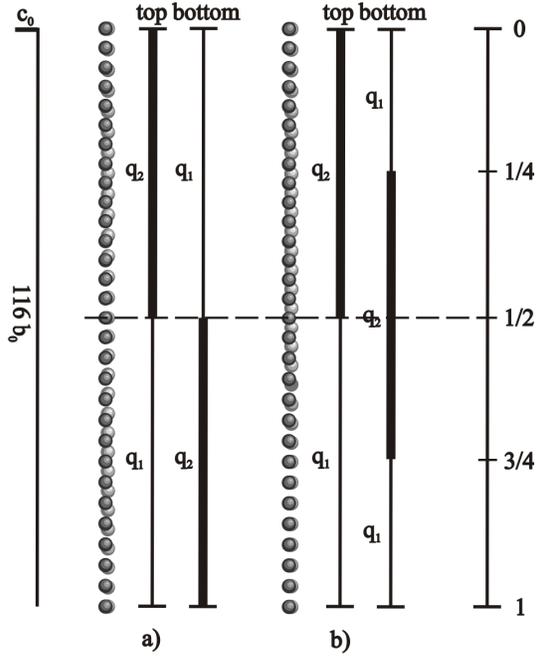

Figure 2 A model of a possible combined $q_1$ and $q_2$ modulation along two adjacent NbSe$_3$ type-III columns (a). The same after the $q_2$ part was displaced along the lower column by π (b).

All details revealed by LT STM, together with the results of the alternative methods, are in accord with a partly modified domain model. According to this, $q_1$ and $q_2$ CDW segments can be interchanged along the type-III columns bellow $T_1$ and in addition along type-I columns bellow $T_2$. The suggested disorder explains the appearance of the CDW domain structure and the characteristic twinkling in the satellite dark field TEM images. The type-II columns remain more or less unmodulated in the entire temperature range between RT and LHeT. Since the type-III pairs of columns form slabs parallel to the $a_0$-$b_0$ plane, the CDW ordering between $T_1$ and $T_2$ also takes place along these slabs. However, contrary to the predictions of the original domain model [43] the LT STM results suggest that the strongly bonded pairs of columns are alternatively modulated by the $q_1$ and $q_2$ CDWs. Accordingly, layered domains of modulated BCTP columns are formed parallel to the $b$-$c$ plane and composed of at least a pair of adjacent $q_1$ and $q_2$ sublayers, which can easily be interchanged as part of the unstable domain structure.

The four modes present bellow $T_2$, i.e. $q_1$ and $q_2$ along both type-III and type-I columns, can under certain conditions be replaced by two only. These require that they appear as coupled pairs, whose composed LP commensurate modulation fits to 58 $b_0$ and stretches laterally across a pair of columns (either type-III or type-I). The IC components of the two wave vectors must in that case add to ½ exactly, which is due to one report [44] and due to a slight variation of the $q_1$ vector with temperature [30] still not proved beyond any doubt.

It appears that the origin of the peculiar CDW ordering in NbSe$_3$ and the structurally related compounds is in its basic structure, composed of symmetry related pairs of BCTP columns, which are alternatively modulated by the two *q* vectors. The rest depends on possible ordering of such units into anisotropic nano-domains with the peculiar contribution to the reciprocal space.

The suggested CDW modulated structure in $NbSe_3$ is in accord with the available experimental evidence and particularly with all details revealed by LT STM and the corresponding Fourier transforms [39,41] and is not in contradiction with the recently reported surface dependence of the $T_2$ transition temperature [42].

## 6. Conclusions

The details revealed by the LT STM studies raise questions regarding a number of aspects of the conventional picture of the CDW states occurring in $NbSe_3$.

The available experimental evidence supports a revised alternative model, based on the existence of coupled layered $q_1$ and $q_2$ nano-domains.

The suggested model brings into accord all important details revealed in the LT STM experiments, like the confinement of the modulation to certain structural segments, the modulation periodicity and its phase relationship with the neighboring columns, and can be brought into accord with the results of a few alternative methods (x-ray, NMR and TEM), which are based on a statistical collection of data.

Further experimental and theoretical studies are needed to clarify these results further and particularly to advance our understanding of the physical reasons for the unique phenomenon in solid state science.

**Acknowledgements**

Financial support of the Slovenian Research Agency (ARRS) (AP, HJPvM, RŽ and EZ) and of the Natural Sciences and Engineering Research Council of Canada (JCB) is gratefully acknowledged.